\documentclass[10pt,twoside]{article}
\usepackage{aas2pp4}
\usepackage[dvips]{graphicx}
\newcommand{\K}{\mbox{$\rm\,K$}} 
\newcommand{\gram}{\mbox{$\rm\,g$}} 
\newcommand{\Msun}{\mbox{$M_\odot$}} 
\newcommand{\second}{\mbox{$\rm\,s$}} 
\newcommand{\yr}{\mbox{$\rm\,yr$}} 
\newcommand{\cm}{\mbox{$\rm\,cm$}} 
\newcommand{\km}{\mbox{$\rm\,km$}} 
\newcommand{\GramPerCc}{\gram\cm^{-3}} 
\newcommand{\erg}{\,{\rm ergs}}	
\newcommand{\MeV}{\mbox{$\rm\,MeV$}} 
\newcommand{\keV}{\mbox{$\rm\,keV$}} 
\newcommand{\gauss}{\mbox{$\rm\,G$}} 
\newcommand{\kB}{k_{\rm B}{}}	
\newcommand{\Mspyr}{\mbox{$\,M_\odot\rm\,yr^{-1}$}}

\newcommand{\ee}[1]{\times 10^{#1}}
\newcommand{\Qnuc}{Q_{\rm nuc}}

\newcommand{\xmm}{{\em XMM\/}}
\newcommand{\axaf}{{\em AXAF\/}}
\newcommand{\rosat}{{\em ROSAT\/}}
\newcommand{\asca}{{\em ASCA\/}}
\newcommand{\beppo}{{\em BeppoSAX\/}}
\newcommand{\rxte}{{\em RXTE\/}}
\newcommand{\kpc}{\mbox{$\rm\,kpc$}}
\newcommand{\timdot}{\langle\dot{M}\rangle}
\newcommand{\tth}{t_{\rm th}}
\newcommand{\Teff}{T_{\rm eff}}
\begin{document}

\markboth{
   Brown, Bildsten, \& Rutledge
}{
   Hot Cores in Transient Neutron Stars
}
\title{
   Crustal Heating and Quiescent Emission from Transiently Accreting
   Neutron Stars  
}
\author{
   Edward F. Brown, Lars Bildsten, and Robert E. Rutledge\altaffilmark{1}
}
\affil{
   Department of Physics and Department of Astronomy\\ 
   601 Campbell Hall, Mail Code 3411, University of California,
   Berkeley, CA 94720--3411;\\ ebrown@astron.berkeley.edu,
   bildsten@fire.berkeley.edu, rutledge@astron.berkeley.edu
}
\authoremail{
   ebrown@astron.berkeley.edu,bildsten@fire.berkeley.edu,
   rutledge@astron.berkeley.edu
}
\altaffiltext{1}{Current Address: Space Radiation Laboratory, California
Institute of Technology, MS 220-47, Pasadena, CA 91125;
rutledge@srl.caltech.edu}

\thispagestyle{empty}
\begin{abstract}

Nuclear reactions occurring at densities $\approx10^{12}\GramPerCc$ in
the crust of a transiently accreting neutron star efficiently maintain
the core at a temperature $\approx (5\mbox{--}10)\ee{7}\K$.  When
accretion halts, the envelope relaxes to a thermal equilibrium set by
the flux from the hot core, as if the neutron star were newly born.  For
the time-averaged accretion rates ($\lesssim 10^{-10}\Mspyr$) typical of
low-mass X-ray transients, standard neutrino cooling is unimportant and
the core thermally reradiates the deposited heat.  The resulting
luminosity is $\sim 5\ee{32}\mbox{--}5\ee{33}\erg\second^{-1}$ and
agrees with many observations of transient neutron stars in quiescence.
Confirmation of this mechanism would strongly constrain rapid neutrino
cooling mechanisms for neutron stars (e.g., a pion condensate).  Thermal
emission had previously been dismissed as a predominant source of
quiescent emission since blackbody spectral fits implied an emitting
area much smaller than a neutron star's surface.  However, as with
thermal emission from radio pulsars, fits with realistic emergent
spectra will imply a substantially larger emitting area.  Other emission
mechanisms, such as accretion or a pulsar shock, can also operate in
quiescence and generate intensity and spectral variations over short
timescales.  Indeed, quiescent accretion may produce gravitationally
redshifted metal photoionization edges in the quiescent spectra
(detectable with \axaf\ and \xmm).  We discuss past observations of
Aql~X-1 and note that the low luminosity ($<10^{34}\erg\second^{-1}$)
X-ray sources in globular clusters and the Be star/X-ray transients are
excellent candidates for future study.

\end{abstract}


\vspace*{2.0cm}
\begin{center}
\Large To appear in \emph{The Astrophysical Journal, part II,} 504
\end{center}

\section{Introduction}
\label{s:Introduction}

The temperature of an accreting neutron star's core and crust is a
subject of much current interest, both for magnetic field evolution
studies and thermonuclear burning (\cite{brown98a} and references
therein).  For neutron stars (NSs) that steadily accrete at $\dot{M}\sim
10^{-11}\mbox{--}10^{-9}\Mspyr$, balancing the recurrent heating from
thermally unstable hydrogen/helium burning with the cooling from
neutrino emission and radiative diffusion requires core temperatures
$T_c=(1\mbox{--}2 )\ee{8}\K$ (\cite{ayasli82}; \cite{hanawa84}).  For
these accretion rates, no more than a few percent of the hydrogen/helium
burning luminosity diffuses inwards and heats the core
(\cite{fujimoto84}, 1987).  There is an additional furnace in the inner
crust (densities above neutron drip, $\sim5\ee{11}\GramPerCc$), where
the compression of matter by accretion induces electron captures,
neutron emissions, and pycnonuclear reactions that release $\Qnuc\approx
1\MeV/m_p\approx 10^{18}\erg\gram^{-1}$ (\cite{haensel90a}).  These
reactions send most of their heat into the core and therefore play a
more important role in maintaining the interior thermal balance of a
constantly accreting neutron star (\cite{miralda-escude90}; 
\cite{bildsten97a}; \cite{brown98a}).

For the case of transient accretors, the heating of the interior is not
so simple.  Most of the heat released by hydrogen/helium burning in the
upper atmosphere leaves immediately during the unstable burning
(\cite{hanawa86}; \cite{fujimoto87}).  Moreover, the time between accretion
outbursts is much longer than the cooling time of the NS atmosphere.  As
a result, H/He burning alone cannot heat the core to the interior
temperatures of a steadily accreting star.  This is an important point,
since thermal emission from transiently accreting NSs would be
observable when accretion halts, if the core were sufficiently hot
(\cite{vanParadijs87}; \cite{verbunt94}; \cite{asai96a}; \cite{campana98b}).

In this Letter, we show that direct nuclear heating in the deep neutron
star crust heats the interior much more strongly and naturally explains
a large fraction of the quiescent emission seen from transient NSs.
During each outburst, the heat released from the crustal reactions flows
into the core and maintains it at $T_c\approx 10^8\K$
(\cite{bildsten97a}).  Once the core is in steady-state, the heat radiated
during quiescence must equal the fraction of the heat deposited during
the outburst.  As we show in \S~2, in quiescence the NS then emits a
time-averaged luminosity $\sim (1\MeV/m_p)\timdot\sim
6\ee{32}\erg\second^{-1} (\timdot/10^{-11}\Mspyr)$.  For time-averaged
(we here mean over the recurrence interval) accretion rates of
$\timdot\lesssim 10^{-10}\Mspyr$, this amount of thermal emission is
unavoidable unless the neutrino cooling is much more rapid than the
standard mechanisms we considered.

Of the NS transients detected in quiescence, most have a luminosity
$L_q\sim 10^{33}\erg\second^{-1}$, and the ratio of outburst to
quiescent fluence is constant to within a factor of $\sim10$
(\cite{chen97}).  The rough agreement between our estimate and observed
values of $L_q$ gives new impetus for a thermal interpretation of the
quiescent emission.  As we discuss in \S~\ref{s:emission}, spectral fits
with realistic emergent spectra will show an increase in the emitting
area over blackbody fits (as in radio pulsars, \cite{rajagopal96};
\cite{zavlin96}).  This alleviates one criticism of the thermal emission
hypothesis, namely that the emitting areas are too small.  Of course,
other luminosity sources, including accretion and a shock from a
rotation-powered pulsar (\cite{campana98b}), may also occur along with
thermal emission.  An important challenge is to identify the luminosity
fraction originating from the NS surface.

\section{Local Heating in the Deep Crust}
\label{s:interior}

Accretion pushes a given fluid element in the crust ever deeper and
forces the electron Fermi energy to grow with time.  The electrons
eventually capture onto nuclei (\cite{haensel90a}), which become
progressively neutron rich and then fuse via pycnonuclear reactions.
This heat is locally deposited in the crust about the thin layers in
which the reactions occur.  Most of the heat is conducted into the core
for a steadily accreting cool NS.  The interior heats up, on a timescale
of $\sim10^4\mbox{--}10^5\yr$, until either the entire crust is nearly
isothermal or, if the accretion rate is sufficiently rapid, until
neutrino cooling balances the heating (Miralda-Escud\'e et~al.\ 1990;
\cite{brown98a}).

\begin{figure}[htb]
\includegraphics[width=\hsize]{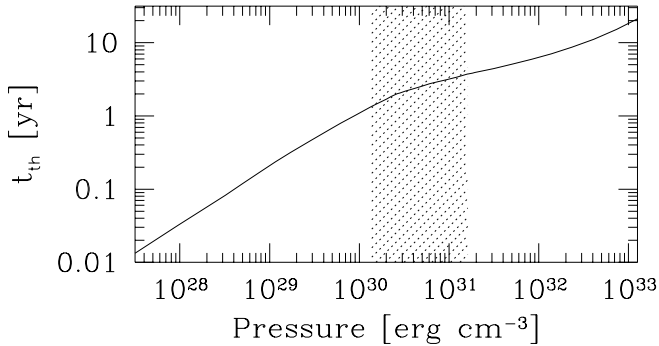}
\caption{Timescale for heat to diffuse from a given pressure to the
surface (eq.\ [\protect\ref{e:thermal}]), for a neutron star ($R=10\km$,
$M=1.4\Msun$) accreting at $\protect\timdot=10^{-11}\Mspyr$.  Most of
the nuclear energy (98\%) is released in the shaded region.  The change
in slope at the shaded region is caused by a reduction in the electron
to baryon ratio.  At $P>10^{31}\erg\cm^{-3}$, the pressure goes from
electron-dominated to neutron-dominated.
\label{f:structure}}
\end{figure}

The situation is different for a transiently accreting NS.  During an
outburst, the reactions locally heat the crust at a rate $\Qnuc\dot{M}$
and raise its temperature by $\sim1\%$.  The layer in which the
reactions occur ({\em shaded region}, Fig.~\ref{f:structure}) is then
hotter than both the atmosphere and the core.  After the rapid accretion
stops, the crust cools and its thermal profile relaxes to that of an
isolated cooling NS, i.e., the temperature monotonically increases with
depth.  While the crust heats and cools in response to the changing
accretion rate, a fraction $f$ of the deposited heat flows inward and
adds to the heat stored in the core, while a fraction $(1-f)$ flows
outward and radiates away.  By equating the sum of the quiescent
luminosity from the cooling core (we use the core temperature-luminosity
relation for an atmosphere with an accreted mass of $3\ee{-11}\,\Msun$;
\cite{potekhin97}) and the neutrino luminosity (modified Urca:
\cite{friman79}; crust neutrino bremsstrahlung: \cite{pethick97})
with the energy deposited during the outburst, we determine $T_c$.  For
example, using the core temperature-luminosity relation for a fully
accreted envelope (Potekhin et~al.\ 1997) implies $T_c\approx 10^8 (f
\langle\dot{M}/10^{-10}\Mspyr\rangle)^{0.4}\K$.  For these temperatures,
neutrino cooling is unimportant.

After many outbursts the core reaches an equilibrium limit cycle, in
which the heat that it loses between outbursts (the {\em incandescent\/}
luminosity $L_i$) is replenished by the fraction $f$ of the nuclear
energy deposited during each outburst.  We expect that all observed
transient NSs will have long since reached this equilibrium limit cycle.
The timescale to heat the core is much shorter than the lifetime of a
low-mass X-ray binary; at $\langle\dot{M}\rangle=10^{-11}\Mspyr$, the
time to heat the core to equilibrium is less than $10^7\yr$ for
$f\gtrsim 0.01$.  Since the energy injected into the crust during the
outburst period $t_o$ reradiates during the recurrence interval $t_r$,
we expect an average incandescent luminosity
\begin{eqnarray}\label{e:Lq}
   L_i &\approx& f\Qnuc\left(\frac{1}{t_r}\int_{\rm
   outburst}\dot{M}\,dt\right) \nonumber\\
   &\equiv& f\Qnuc\langle\dot{M}\rangle.
\end{eqnarray}
This relation, although simple, is not useful observationally because
both sides of the equation depend on the source distance.  To divide out
this dependence, we rewrite equation (\ref{e:Lq}) in terms of fluences.
The outburst fluence is $L_ot_o=t_r\langle\dot{M}\rangle GM/R$ and
the incandescent fluence is $L_it_r$.  The ratio of the fluences is then
simply
\begin{equation}\label{e:ratio}
   \frac{L_ot_o}{L_it_r}=\frac{GM/R}{f\Qnuc}\approx \frac{200}{f}.
\end{equation}
Because the core temperature cannot appreciably change in a single
outburst, there will be little change in the incandescent luminosity
from one outburst cycle to the next.

While the crust relaxes to a cooling thermal profile and radiates the
fraction $1-f$ of the deposited energy, the neutron star luminosity
asymptotically approaches $L_i$ on the thermal diffusion timescale
(\cite{henyey69}),
\begin{eqnarray}
\label{e:thermal}
\tth &=& \frac{1}{4} \left[\int_0^P \left(\frac{c_P}{\rho
   K}\right)^{1/2} \frac{dP}{g} \right]^2 \nonumber\\ &\approx& 1\yr
   \left(\frac{2\ee{14}\cm\second^{-2}}{g}\right)^2
   \left(\frac{P}{10^{31}\erg\cm^{-3}}\right)^{3/4}.
\end{eqnarray}
Here $P$ is the pressure, $g$ is the surface gravity, $c_P$ is the
specific heat, and $K$ is the conductivity (see \cite{brown98a} for
details).  In Figure \ref{f:structure} we plot $\tth$ as a function of
pressure for a NS accreting at $\timdot=10^{-11}\Mspyr$.  The thermal
diffusion time in the crust is insensitive to the temperature and hence
is almost independent of accretion rate.

For objects with lower outburst fluences (e.g., Aql~X-1, for which
$t_r\approx 1\yr$ and $t_o\approx 30$~days), most of the heat is stored
in the core, i.e., $f\approx 1$.  The reason is that the energy
deposited in the outburst only raises the crust temperature by
$\sim0.1\,\langle\dot{M}/10^{-11}\Mspyr\rangle(t_r/1\yr)\%$.  As a
result, the luminosity radiated as the crust relaxes is only $\sim
0.3\%$ different from the incandescent luminosity.  Note that this
argument does not hold for objects with similar $\langle\dot{M}\rangle$
but longer recurrence times, such as Cen~X-4 and 4U~1608--52.  For these
objects, the crust temperature can change by $\gtrsim10\%$ during an
outburst, and so we expect $f<1$.

\begin{figure}[htb]
\includegraphics[width=\hsize]{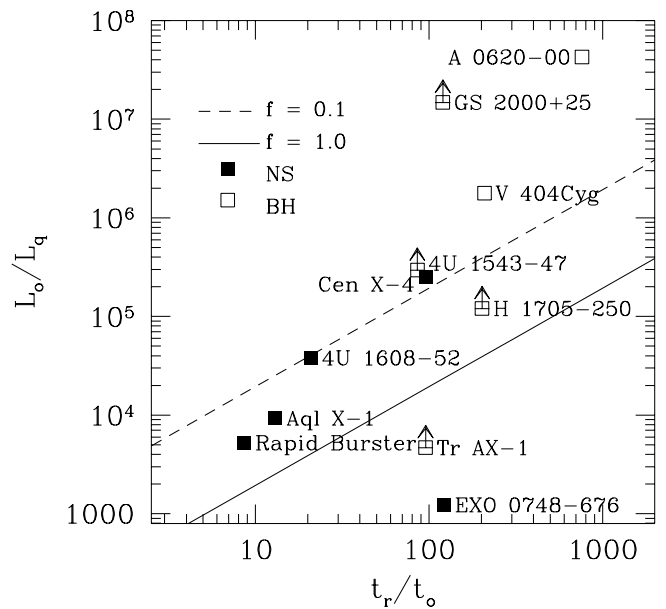}
\caption{The ratio of outburst luminosity $L_o$ to quiescent luminosity
$L_q$ as a function of the ratio of recurrence interval $t_r$ to
outburst duration $t_o$.  The lines are for different fractions, $f=0.1$
({\em dashed line\/}) and $f=1.0$ ({\em solid line\/}), of
$\Qnuc=1.0\MeV$~per baryon mass deposited at depths where the thermal
time is longer than the outburst recurrence time.  Also plotted are the
observed ratios for several NSs ({\em filled squares}) and BHs ({\em
open squares}).  For most of the BHs, only an upper limit ({\em
arrow\/}) to $L_q$ is known.  Data is from Chen et~al.\ (1997), with the
exception of $L_q$ for the Rapid Burster (\protect\cite{asai96b}).  For
Aql~X-1 and the Rapid Burster, $L_o$ and $t_o$ are accurately known
(\rxte/All-Sky Monitor public data); for the remaining sources $L_o$ and
$t_o$ are estimated from the peak luminosities and the rise and decay
timescales.
\label{f:luminosity}}
\end{figure}

Figure \ref{f:luminosity} displays $L_o/L_q$ ($L_q$ is the observed
quiescent luminosity) as a function of $t_r/t_o$ for some NS ({\em
filled squares\/}) and black hole (BH) ({\em open squares\/})
transients.  Note that the BH candidates appear uncorrelated compared to
the NSs.  We also plot the relation for $L_i$ (eq.~[\ref{e:ratio}]) for
different values of $f$.  EXO~0748--676 has $L_o/L_q\approx 1000$ and
from the diagram appears likely to accrete during quiescence. Indeed, it
is presently accreting (as indicated by the \rxte/All-Sky Monitor) at an
even higher rate ($L\sim 10^{36}\erg\second^{-1}$) than when the $L_q$
for Figure \ref{f:luminosity} was measured.  Repeated observations after
the NS goes into quiescence can discern the fraction of the heat
radiated as the crust relaxes, i.e., $1-f$.  The sum of this fluence and
the incandescent fluence is just $\Qnuc\timdot t_r$.  These observations
can in principle constrain $g$, on which the thermal time strongly
depends (eq.~[\ref{e:thermal}]).

\section{The Repercussions of Deep Heating}\label{s:emission}

\subsection{Observational Signatures}

The expected heating in the crust naturally generates an incandescent
luminosity $L_i$ that matches the observed values of $L_q$.  Of course,
the star may still accrete, either persistently or spasmodically, in
quiescence (\cite{narayan97a}).  Accretion onto the NS surface at low
rates will also produce a thermal spectrum (\cite{zampieri95}), so a
further challenge is to distinguish between the two luminosity sources.
If the NS does not accrete during quiescence, there will be (1) little
variation in $L_q$ from one outburst cycle to the next (we here mean the
incandescent luminosity emitted after the crust has relaxed to a cooling
thermal configuration), (2) a slow monotonic decrease in flux and
$T_{\rm eff}$ while the crust relaxes after the outburst, (3) no
short-term intensity fluctuations (barring environmental effects), (4)
stability of the ratio of the NS radius to distance (as inferred from
spectral fitting), and (5) an absence of metals in the spectra.

The quiescent energy spectra of 1608--52, Cen~X-4 (both observed with
\asca; \cite{asai96a}), and Aql~X-1 (observed with \rosat/Position
Sensitive Proportional Counter; \cite{verbunt94}), when fit by a
blackbody (BB) with an additional power-law tail for Cen~X-4, imply
substantially smaller radii ($1.5\km$ for 1608--52, $1.8\km$ for
Cen~X-4, and $1\km$ for Aql~X-1) than expected from a NS for the
best-fit BB temperatures (0.16--0.30\keV).  These measured radii have
puzzled observers and motivated the interpretation that either the
quiescent NS luminosity does not originate from the surface or only some
small fraction of the surface radiates.  {\em However, the emergent
spectra at $\Teff\lesssim 5\ee{6}\K$ is very different from a BB.}  The
opacity is free-free dominated and is therefore proportional to
$\nu^{-3}$, where $\nu$ is the photon frequency.  As a result, higher
energy photons escape from greater depths, where $T>\Teff$
(\cite{romani87}; \cite{zampieri95}).  Spectral fits of the Wien tail with BB
curves then overestimate $T_{\rm eff}$ and underestimate the emitting
area, by as much as orders of magnitude (\cite{rajagopal96};
Zavlin et ~al.\ 1996).  Simple comparisons between the observed
spectra and the hydrogen atmosphere models imply larger emitting areas
in all cases.

An incandescent NS should have a pure H atmosphere because the time for
heavy elements to settle out of the photosphere is $\sim 10\second$
(\cite{romani87}). However, if accretion continues in quiescence, there
is an accretion rate, $\dot{M}_{\rm Z}$, above which metals are dumped
into the atmosphere fast enough to maintain their abundance.  {\em In
this case a measurable metal abundance may exist.}  In the absence of
spallation (which depends on the accretion geometry), metals will be
underabundant relative to their infalling value for
$\dot{M}<\dot{M}_{\rm Z} \approx 4\times 10^{-14}\Mspyr (\kB T_{\rm
eff}/0.1\keV)^{3/2} (8/Z)$ (\cite{bildsten92}).  The accretion rate
$\dot{M}_{\rm Z}$ coincides with the accretion rate at which the
resulting luminosity from quiescent accretion, $GM\dot{M}/R$, is
comparable to the thermal emission.  {\em If accretion alters the
luminosity, it also alters the photospheric abundances.}  The emergent
spectra from an atmosphere with solar abundance metals is detectably
different from that of a pure H atmosphere (\cite{rajagopal96}), mostly
because of the gravitationally redshifted O~{\sc viii} photoionization
edge.  This feature (at $0.87\keV$ in the rest frame of the ion), if
detectable with \xmm, would allow direct measurement of the
gravitational redshift.

Thermal emission does not explain the hard power-law tail seen in both
Cen~X-4 (\cite{asai96a}) and Aql~X-1 (\cite{campana98a}).  Evidence of an
additional luminosity source is also indicated by the fast (timescales
of order days) variations in the quiescent flux observed from Cen~X-4
(\cite{campana97}).  If a rotation-powered pulsar becomes operational in
quiescence, the pulsar wind can interact with material in the NS
environment, similar to that in the pulsar/Be star system PSR~1259--63
(\cite{tavani97}).  The emission from the shock is most likely nonthermal,
however, so future observations can distinguish the part of the spectrum
contributed by thermal emission.

\subsection{Aql~X-1 and SAX~J1808.4--3658}

Recent observations of Aql~X-1 as it faded into quiescence
(\cite{campana98a}; \cite{zhang98b}) revealed an abrupt (1~day) decay of the
luminosity following a more gradual decline with a roughly 17~day
timescale.  After the sudden dimming, the luminosity persisted at
$L_q\approx10^{33}\erg\second^{-1}$ for the remainder of the
observation.  Both Campana et~al.\ (1998b) and Zhang et~al.\ (1998)
interpreted this sharp transition as the onset of the propeller effect
(centrifugal barrier).  Accretion at the low rates ($\approx
10^{-13}\Mspyr$) needed to explain $L_q$ is very difficult if the
propeller is operational (\cite{stella94}).  In addition, accretion in
spite of the propeller implies a flow onto the polar caps that would
produce luminosity variations at the NS spin period, which are not seen.
The propeller can be avoided only if the NS in Aql~X-1 rotate with
$P_s>0.6(B/10^9 \gauss)^{6/7}\second$ (\cite{verbunt94}).  Interpreting
the $\rm549\,Hz$ oscillation seen during a type I burst from Aql~X-1
(\cite{zhang98a}) as the rotational frequency would then imply that
$B<10^6\gauss$.  This magnetic field constraint relaxes substantially if
thermal emission causes $L_q$ and makes more plausible the NS becoming
an active millisecond radio pulsar.

The steadiness of $L_q$ (over a 20~day observation with \beppo,
\cite{campana98a}) is naturally explained by thermal emission,
provided that $f\approx 1$ (which we expect from the arguments in
\S~2).  The incandescent emission from the hot NS has the correct
magnitude to explain the observations without requiring any ad
hoc assumptions.  Alternate explanations, such as accretion onto the
magnetopause or shock emission from a rotation-powered pulsar
(\cite{campana98b}), are unlikely to produce both a steady luminosity and
a thermal spectrum, especially one with an emitting radius comparable to
that of a NS.  However, at least one other emission mechanism is
required to account for the hard tail.

With the recent discovery of a $401\rm\,Hz$ accreting pulsar
(\cite{chakrabarty98}; \cite{wijnands98}) in the transient SAX~J1808.4--3658
(\cite{intZand98}), an upper bound of $6\ee{26}\gauss\cm^{3}$ may be
placed on the dipole magnetic moment if the magnetospheric radius is
presumed to be less than the corotation radius.  Given a recurrence
interval of $1.5\yr$, an outburst duration of $\approx20$~days, and an
outburst accretion rate of $\approx3\ee{-10}\Mspyr$ (\cite{intZand98}), we
expect an incandescent luminosity of $\approx
6\ee{32}\erg\second^{-1}(f/1.0)$, which corresponds to an unabsorbed
flux ($4\kpc$ distance) of $3\ee{-13}\erg\cm^{-2}\second^{-1}(f/1.0)$.
The inferred surface magnetic field and spin period are sufficient to
power a rotation-powered pulsar. If this occurs, the magnetospheric
emission is $L_x< 4\ee{32}\erg\second^{-1}$ (\cite{becker97}) and may
contaminate the thermal emission.

\section{Conclusions}\label{s:sources}

We have demonstrated that reactions in the inner crust of an accreting
NS heat the interior enough to make the NS incandescent after accretion
halts.  Transiently accreting NSs then radiate $\sim
5\ee{32}\mbox{--}5\ee{33}\erg\second^{-1}$ in quiescence.  The quiescent
emission is a thermometer that probes the heated deep crust and core.
As in the case of isolated cooling NSs (see, e.g., \cite{tsuruta98}),
it may be possible to test for the presence of enhanced neutrino
emissivities by placing limits on the interior temperature of the NS.

Upcoming missions will greatly increase our understanding of previously
discovered transients and open up new populations for study.  Especially
promising are the low-luminosity X-ray sources in globular clusters
(\cite{hertz83b}).  A subject of debate is whether the brighter sources
of this class (with $10^{33}\erg\second^{-1} < L_x <
10^{34.5}\erg\second^{-1}$) are quiescent soft X-ray transients, as
opposed to cataclysmic variables (\cite{verbunt84}).  If so, future
X-ray observations can mine a potentially rich source of NS
transients. With multiple quiescent NSs in a single field of view, all
at the same distance and reddening, comparisons between sources will be
easier and radius determinations more certain.  Although in many ways
distinct from the low-mass systems, the neutron stars in Be transients
(\cite{apparao94}; \cite{vandenHeuvel87}) should also behave as we have
discussed.  These NSs have strong magnetic fields
($B\gtrsim10^{11}\gauss$), show pulsations, and undergo a variety of
outbursts (\cite{bildsten97b}).

\acknowledgements

We thank G. Pavlov and S. Zavlin for useful discussions, the referee for
comments that substantially improved this paper, and S. Robertson for
alerting us to an error in the preprint.  Our research was supported by
NASA via grants NAG 5-2819 and NAGW-4517.  E.~F.~B. was supported by a
NASA GSRP Graduate Fellowship under grant NGT-51662.  L.~B. acknowledges
support as an Alfred P. Sloan Foundation fellow.

\normalsize

\end{document}